\documentclass[twocolumn,showpacs,amssymb,nobibnotes,aps,floats,psfig,prb]{revtex4-1}
\usepackage{textcomp,amssymb,graphicx,epsf}
\usepackage{hyperref}
\usepackage{amsbsy}
\usepackage{amsmath}
\usepackage[dvips]{color}
\usepackage{graphicx,epsfig}
\usepackage{epsfig}
\usepackage{float}
\date{}
\def \s{\sigma}
\def \sb{\bar\sigma}
\def \be{\begin{eqnarray}}
\def \ee{\end{eqnarray}}

\def \d{\dagger}
\def \sd{\sum\limits}
\def \sp{\!\!\!}
\def \spb{\sp\sp}
\begin{document}
\title{Phase diagram of one-dimensional Hubbard-Holstein model at quarter-filling}
\author{Sahinur Reja${^1}$}
\author{Sudhakar Yarlagadda${^1}$}
\author{Peter B. Littlewood${^2}$}
\affiliation{${^1}$CAMCS and TCMP Div., Saha Institute of Nuclear Physics,
Kolkata, India}
\affiliation{${^2}$Cavendish Lab, Univ. of Cambridge, UK}
\date{\today}
\begin{abstract}
We  derive an effective Hamiltonian for the one-dimensional Hubbard-Holstein model, valid
in a regime of both strong electron-electron (e-e) and electron-phonon (e-ph) interactions and in the non-adiabatic limit
($t/\omega_0 \leq 1$), by using a non-perturbative approach.
We  obtain the phase diagram at quarter-filling
by employing a modified Lanczos method and studying various density-density correlations.
The spin-spin AF (antiferromagnetic) interactions and nearest-neighbor repulsion, resulting from the e-e and the e-ph interactions
respectively, are the dominant terms (compared to hopping) and compete to determine the various correlated phases.
As e-e interaction $(U/t)$ is increased, the system
 transits from an AF cluster to a correlated singlet phase through a discontinuous transition at all
strong e-ph couplings $2 \leq g \leq 3$ considered. At higher
values of
$U/t$ and moderately strong e-ph interactions ($2 \leq g \leq 2.6$),
the singlets break up to form an AF order and then to a paramagnetic order all in a single sublattice; whereas
at larger values of $g$ ($> 2.6$), the system jumps directly to the spin disordered charge-density-wave (CDW) phase.
\end{abstract}
\maketitle
\section{Introduction:}
A host of materials show evidence of e-ph interactions besides the expected e-e interactions.
Angle-resolved photoemission spectroscopy (ARPES) experiments in cuprates
\cite{photoem1,photoem3}, fullerides \cite{fullarene3}, and  manganites \cite{lanzara2}
indicate strong e-ph coupling.
The interplay of e-e and e-ph interactions in these correlated systems
leads to  coexistence of or competition between various phases
such as superconductivity,  CDW, spin-density-wave (SDW)  phases, or formation of novel non-Fermi liquid phases,
 polarons, bipolarons, etc. It is of particular interest to consider both the strong coupling regime, 
and the cross-over to the non-adiabatic limit where the lattice response is not slower
in comparison to the heavy effective mass of the correlated electron system.

The simplest framework to analyze the effects of e-e and the concomitant e-ph interactions
is offered by the Hubbard-Holstein model whose Hamiltonian is given by:
\be
H_{hh} \!&=&\! -t\sum_{j\sigma}\left(c^{\dagger}_{j+1\sigma}c_{j\sigma}+ {\rm H.c.} \right)
+\omega_0\sum_{j}a_{j}^{\dagger}a_{j}\nonumber\\
\spb && +g\omega_0\sum_{j\sigma}n_{j\sigma}(a_{j}+a_{j}^{\dagger})
+U\sum_{j}n_{j\uparrow}n_{j\downarrow}  .
\label{ai1}
\ee
Here,  $n_{j\sigma} \equiv c_{j\sigma}^{\dagger}c_{j\sigma}$ with $c_{j\sigma}^{\dagger}$ and $a_{j}^{\dagger}$ being
the creation operators at site $j$ for  an electron with spin
$\sigma$ and a phonon respectively.
The Hamiltonian describes a tight-binding model with hopping amplitude $t$, a set of independent oscillators characterized
by a dispersionless phonon frequency $\omega_0$, along with an onsite Coulomb repulsion of strength $U$ and an onsite
electron-phonon interaction of strength $g$.
The Holstein and the Hubbard models are recovered in the limits $U=0$ and $g=0$ respectively.

To gain insight into the rich physics of the Hubbard-Holstein model,
several studies have been conducted
(in one-, two-, and infinite-dimensions and at various fillings)
by employing various approaches such as
quantum Monte Carlo (QMC) \cite{qmc1,qmc3,qmc4,qmc5,qmc6,qmc7},
exact diagonalization \cite{exdiag1,exdiag2,exdiag4},
density matrix renormalization group (DMRG)\cite{dmrg},
dynamical mean field theory (DMFT)
\cite{dmft1,dmft2,dmft3,dmft4,dmft5,dmft6,dmft7,dmft8},
semi-analytical slave
boson approximations \cite{slave_b1,slave_b2,slave_b3},
 large-N expansion \cite{largeN2}, variational methods
based on Lang-Firsov transformation \cite{LF}, and  Gutzwiller
approximation
\cite{GA}.

In contrast to earlier approaches,
we utilize a
controlled analytic approach (that takes into account quantum phonons).  
Our method uses both the strong electron-phonon coupling limit $g > 1$ and the 
strong Coulomb coupling limit $U/t > 1$ to generate an effective $t-J$ model with 
displaced oscillators that can then be treated perturbatively, and generates 
longer-range interactions in an effective Hamiltonian. This model we then solve 
numerically for finite chains.
We then obtain
the phase diagram of the Hubbard-Holstein model at quarter-filling in one dimension.

Our effective Hamiltonian comprises
of two dominant competing interactions -- spin-spin AF interaction and nearest-neighbor (NN) electron repulsion.
In addition, three types of hopping also result -- the NN hopping with reduced band width, next-nearest-neighbor (NNN)
hopping, and NN spin-pair $\s \sb$ hopping.
As the  e-e interaction $U/t$ is increased, the system sequentially transforms from an AF cluster phase to a correlated
singlet phase followed by CDW phase(s) with (e-ph coupling $g$ dependent) accompanying spin order.
The most  interesting feature
is that, at intermediate
values of $U/t$ and for all strong e-ph couplings ($2 \leq g \leq 3$) considered, a phase comprising of correlated
NN singlets appears which suggests the possibility for superconductivity occurrence.

The paper is organized as fallows. In section II, we describe our non-perturbative
approach for tackling strong e-ph interaction and derive the effective Hamiltonian;
in section III, we discuss
the region of parameter space for $U/t$, $\omega_0/t$, and $g$ where our derived results are applicable; next, different
correlation functions are analyzed
to obtain the various phases and the conditions for transitions between these phases are presented in section IV; then,
in section V, we present
the phase diagram; and in section VI our conclusions.

\section{Effective Hamiltonian}
To get the effective Hubbard-Holstein Hamiltonian, we first carry out the
well-known Lang-Firsov (LF) transformation \cite{lang}
$H^{LF}_{hh}=e^{T}H_{hh}e^{-T}$
where $T=-g\sum_{j\sigma}n_{j\sigma}(a_{j}-a_{j}^{\dagger})$.
This transformation clothes the hopping electrons with phonons and
displaces the simple harmonic oscillators.
On rearranging the various terms, we get the following LF transformed
Hamiltonian: \\
\be
H^{LF}_{hh}&=& -t\sum_{j\sigma}(X_{j+1}^{\dagger}c_{j+1\sigma}^{\dagger}c_{j\sigma}X_{j}+ {\rm H.c.})+\omega_{0}\sum_{j}
a_{j}^{\dagger}a_{j}\nonumber\\
&& -g^{2}\omega_{0}\sum_{j}n_{j}+(U-2g^2\omega_{0})\sum_{j}n_{j\uparrow}n_{j\downarrow} ,
\label{ai2}
\ee
where $X_{j}=e^{g(a_{j}-a_{j}^{\dagger})}$ and $n_j = n_{j \uparrow} + n_{j \downarrow}$.
On noting that $X_{j}^{\dagger}X_{j}=1$,
we can express, as shown below,
 our LF transformed Hamiltonian in terms of the composite
fermionic operator
$d_{j\sigma}^{\dagger} \equiv c_{j\sigma}^{\dagger}X_{j}^{\dagger}$
(i.e., a fermionic operator dressed with phonons):
\be
H^{LF}_{hh}= -t\sum_{j\sigma}\left(d_{j+1\sigma}^{\dagger}d_{j\sigma}+
{\rm H.c.}\right)+\omega_{0}\sum_{j}
a_{j}^{\dagger}a_{j}\nonumber\\
+(U-2g^2\omega_{0})\sum_{j}
n_{j\uparrow}^{d}n_{j\downarrow}^{d}
-g^{2}\omega_{0}\sum_{j}\left(n_{j\uparrow}^{d}+n_{j\downarrow}^{d}\right) ,
\label{ai3}
\ee
where
 $ n_{j\sigma}^{d}= d_{j\sigma}^{\dagger}d_{j\sigma} $.
The last term in Eq. (\ref{ai3}) represents the
polaronic energy and is a constant for a given number of particles
and an associated set of parameters.
Hence,  we drop this term from now onwards.

 The model represented by Eq. (\ref{ai3}) is essentially
the Hubbard Model for composite fermions
 with Hubbard interaction $U_{eff}=(U-2g^2\omega_0)$
and can be directly converted to an effective $t-J$ model Hamiltonian
in the limit of large $U_{eff}/t$.
 Thus, we get the effective $t-J$ Hamiltonian for composite fermions
by projecting out double occupation.
\be
H_{t-J}&=&P_s\left[ -t\sum_{j\sigma}\left(d_{j+1\sigma}^{\dagger}d_{j\sigma}+{\rm H.c.}\right)
+\omega_{0}\sum_{j}a_{j}^{\dagger}a_{j} \right. \nonumber\\
&+& \left. J\sum_{j}\left(\vec{S}_{j}\cdot\vec{S}_{j+1}-{n_{j}^{d}n_{j+1}^{d}\over{4}}\right)\right]P_s ,
\label{ai5}
\ee
where $n_j^{d} =  n_{j\uparrow}^{d}+n_{j\downarrow}^{d}$,
$J={4t^2\over{U-2g^2\omega_{0}}}$, $\vec{S}_i$ is the quantum mechanical spin operator for a spin $1/2$
fermion at site $i$, and $P_{s}$ is the operator
that projects
onto the singly occupied subspace.
Next, we note that
\be
X_{j+1}^{\dagger}X_{j} &=& e^{g(a_{j+1}^{\dagger}
-a_{j+1})}e^{g(a_{j}-a_{j}^{\dagger})}
\nonumber \\
&=&
e^{-g^2}e^{g(a_{j+1}^{\dagger}-a_{j}^{\dagger})}e^{-g(a_{j+1}-a_{j})} .
\label{ai6}
\ee

 The effective $t-J$ Hamiltonian, given in Eq. (\ref{ai5}), can be re-expressed in terms of fermionic
 operators as
\be
H_{t-J}=H_{0}+H_{1} ,
\label{ai7.0}
\ee
where
\be
H_{0}&=& -te^{-g^2}\sum_{j\sigma}P_s \left(c_{j+1\sigma}^{\dagger}c_{j\sigma}+ {\rm H.c.} \right)P_s
+\omega_{0}\sum_{j}a_{j}^{\dagger}a_{j}\nonumber\\
&&+J\sum_{j}P_s\left(\vec{S}_{j}\cdot\vec{S}_{j+1}-{n_{j}n_{j+1}\over{4}}\right)P_s
\label{ai7.1} ,
\ee
and
\be
\!\!\!\!\!H_{1}&=& -te^{-g^2}\!\sum_{j\sigma}P_s \!\left[c_{j+1\sigma}^{\dagger}c_{j\sigma}(Y_{+}^{j\dagger}
Y_{-}^{j}-1)+ {\rm H.c.}\right]\!P_s .
\label{ai7.2}
\ee
Here, we have separated the Hamiltonian into (i) an itinerant electronic system represented by $H_{0}$
containing nearest-neighbor hopping with a reduced amplitude ($t e^{-g^2}$), electronic
 interactions, and no electron-phonon interaction; and (ii) the remaining part $H_1$ which is a perturbation
and corresponds
to the composite fermion terms containing the e-ph interaction with
$Y^{j}_{\pm} \equiv e^{\pm g(a_{j+1}-a_{j})}$.

\subsection{Perturbation Theory}
The unperturbed Hamiltonian $H_{0}$ is characterized by the eigenstates
$|n,m\rangle\equiv|n\rangle_{el}\otimes|m\rangle_{ph}$ and corresponding
 eigenenergies $E_{n,m}=E_{n}^{el}+E_{m}^{ph}.$
On noting that the first-order perturbation term is zero (i.e., $\langle0,0|H_{1}|0,0\rangle=0$),
we proceed to calculate the second-order perturbation term
$E^{(2)}=\sd_{n,m}{{\langle0,0|H_{1}|n,m\rangle\langle n,m|H_{1}|0,0\rangle}\over{E_{0,0}-E_{n,m}}}$
in a manner similar to
that introduced in Ref. \onlinecite{sdadys}.
Now, $\Delta E_{m}=E_{m}^{ph}-E_{0}^{ph}$ is a positive integral multiple of $\omega_{0}$  and $E_{n}^{el}-E_{0}^{el}
\sim te^{-g^2}\equiv t_{eff}$. In the narrow band limit, that $t_{eff}\ll \omega_{0} \ll U_{eff}$,
we get the corresponding second-order perturbation term in the effective Hamiltonian for the polarons to be (see Appendix A
for details):\\
\be
H^{(2)}=\sd_{m}{{_{ph}\!\langle0|H_{1}|m\rangle_{ph}\times{_{ph}\!\langle m|H_{1}|0\rangle_{ph}}}\over
{-\Delta E_m}} .
\label{ai8}
\ee
Evaluation of $H^{(2)}$ leads to
the following expression:
\be
H^{(2)}&=&\sd_{j\s\s^{'}}P_s\left[
V
 \left(c_{j+1\s}^{\dagger}c_{j\s}c_{j\s^{'}}^{\dagger}
c_{j+1\s^{'}} + c_{j\s}^{\dagger}c_{j+1\s}c_{j+1\s^{'}}^{\dagger}c_{j\s^{'}}\right)
\right.\nonumber\\
&+&\! 
t_2
 \left(c_{j-1\s}^{\dagger}c_{j\s}c_{j\s^{'}}^{\dagger}c_{j+1\s^{'}} + c_
{j+1\s}^{\dagger}c_{j\s}c_{j\s^{'}}^{\dagger}c_{j-1\s^{'}} \right. \nonumber\\
&+& \left.\left. c_{j\s}^{\dagger}c_{j+1\s}c_{j-1\s^{'}}^{\dagger}c_{j\s^{'}} +
c_{j\s}^{\dagger}c_{j-1\s}c_{j+1\s^{'}}^{\dagger}c_{j\s^{'}} \right) \right]P_s ,
\label{ai9}
\ee
where $V= (2f_1+ f_2)t_{eff}^2/\omega_{0}$ and $t_2=f_1 t_{eff}^2/\omega_{0}$ 
with
 $f_1=\sd_{n=1}^\infty\frac{g^{2n}}{n!n}$ and $f_2=\sd_{n,m=1}^\infty\frac{g^
{2(m+n)}}{n!m!(m+n)}$. The value of $f_1$ and $2f_1+f_2$ can be approximated for 
large value of $g$ as $e^{g^2}/g^2$ and $e^{2g^2}/{2g^2}$ respectively. But in
numerical simulation we have calculated the value of $f_1$ and $f_2$
by summing over the actual series.  

Finally, to calculate each term in Eq. (\ref{ai9}), we must project out the
double occupancy. We do this projection by replacing every fermionic operator $c_{j\sigma}$ with
the fermionic operator $ c_{j\sigma} (1-n_{j\sb})$ (see Appendix B for details).
\subsection{Effective Electronic Hamiltonian}
\begin{figure}[]
\includegraphics[width=0.98\linewidth]{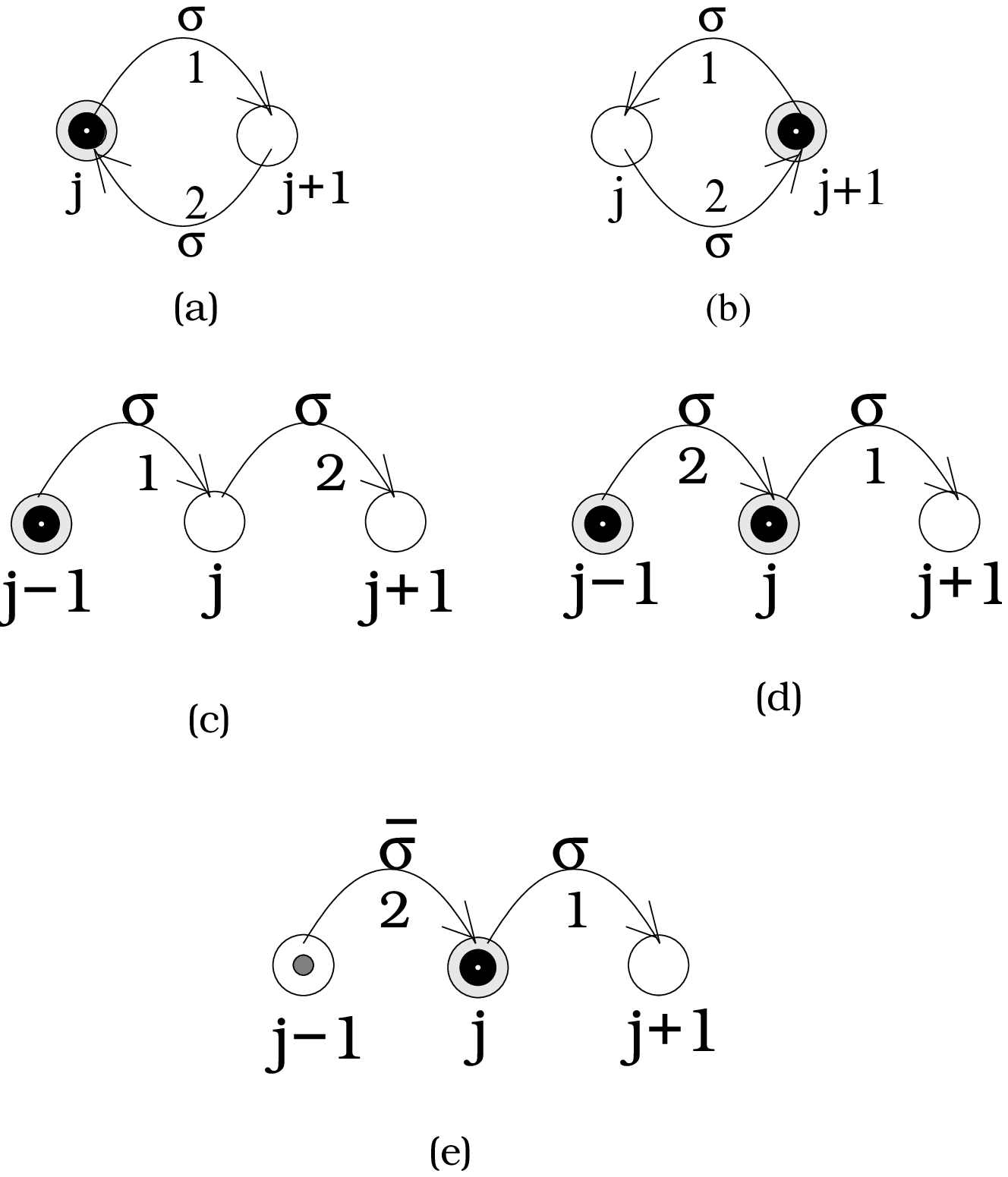}
\caption{{Different hopping processes contributing to second order perturbation
theory are: (a) $c_{j\s}^{\d}c_{j+1\s}c_{j+1\s}^{\d}c_{j\s}$;
 (b) $c_{j+1\s}^{\d}c_{j\s}c_{j\s}^{\d}c_{j+1\s}$;
(c) $c_{j+1\s}^{\d}c_{j\s}c_{j\s}^{\d}c_{j-1\s}$;
(d) $c_{j\s}^{\d}c_{j-1\s}c_{j+1\s}^{\d}c_{j\s}$; and
(e) $c_{j\sb}^{\d}c_{j-1\sb}c_{j+1\s}^{\d}c_{j\s}$.
Here, empty circles correspond to sites with no electrons while
 circles with big and small dots inside correspond to
 sites with spin $\s$ and spin $\sb$ electrons respectively. 
The numbers  $1$ and $2$ indicate the order of hopping. }
}
\label{hop_fig}
\end{figure}

{
The effective Hamiltonian, after averaging over the phononic degrees of freedom, is given by
\begin{widetext}
\be
H_{hh}^{eff}&\cong&-t_{eff}\sd_{j\sigma}P_s \left(c_{j+1\sigma}^{\dagger}c_{j\sigma}+ {\rm H.c.}\right)P_s
+J\sd_{j}P_s \left
(\vec{S}_{j} \cdot \vec{S}_{j+1}-\frac{1}{4}n_{j}n_{j+1}\right)P_s \nonumber \\
&& - \frac{t^2}{2g^2\omega_{0}}
\sd_{j\s}(1- \!\!n_{j+1\sb})(1- \!n_{j\sb})(n_{j\s}-
n_{j+1\s})^2\nonumber\\
&&-\frac{t^2e^{-g^2}}{g^2\omega_{0}} \sd_{j\s}(1-n_{j+1\sb})(1-n_{j\sb})(1-n_{j-1\sb})\left[c_{j+1\s}^\d(1-2n_{j\s})
c_{j-1\s}+ {\rm H.c.} \right]\nonumber\\
&&-\frac{t^2 e^{-g^2}}{g^2\omega_{0}}\sd_{j\s}(1-n_{j+1\sb})(1-n_{j-1\s})
\left[c_{j\s}^{\d}c_{j+1\s}c_{j-1\sb}^\d c_{j\sb} + {\rm H.c.} \right] .
\label{ai10}
\ee
\end{widetext}
Here, we have approximated the coefficients as $V\simeq t^{2}/{2g^2\omega_0}$ and
 $t_2\simeq t^2 e^{-g^2}/{g^2\omega_0}$
 which is valid for large values of $g$.
 The operator form of the last three terms in Eq. (\ref{ai10}) can be visualized
by considering different hopping processes depicted in Fig. \ref{hop_fig}.
The third term 
in Eq. (\ref{ai10}) depicts the process
where an electron hops to its neighboring
site [see Figs. \ref{hop_fig}(a) and \ref{hop_fig}(b)] and returns back. These
two processes add up to the term $(n_{j\s}-n_{j+1\s})^2$ whose expansion yields
the NN repulsion term $2 n_{j\s}n_{j+1\s}$.  
The fourth term is composed of the two hopping processes shown in 
Figs. \ref{hop_fig}(c) and \ref{hop_fig}(d). Fig. \ref{hop_fig}(c) represents
a double hopping process where a spin $\sigma$ particle hops first to its
NN site and then to its NNN site. Contrastingly, Fig. \ref{hop_fig}(d)
depicts the sequential process involving a pair of spin $\s$ electrons
where an electron at site 
$j$ hops to its neighboring site
 $j+1$ followed by
 another 
electron at site $j-1$ hopping to site $j$. This sequential
process  may be called $\sigma\s$ pair hopping. The fifth term corresponds to
Fig. \ref{hop_fig}(e) and represents a hopping process similar to that shown
in Fig. \ref{hop_fig}(d) but involving a pair of electrons with opposite spins $\sb\s$. 
The additional factors
$(1-n_{j+1\sb})(1-n_{j\sb})$, $(1-n_{j+1\sb})(1-n_{j\sb})(1-n_{j-1\sb})$,
and $(1-n_{j+1\sb})(1-n_{j-1\s})$
appearing respectively in the third, fourth, and fifth terms in Eq. (\ref{ai10})
represent projecting out double-occupancy at a site.
The detailed derivation of these projected terms is given in 
Appendix \ref{proj_singly}.
}

\begin{figure}[htb]
\includegraphics[height=4in,width=3in,angle=0]{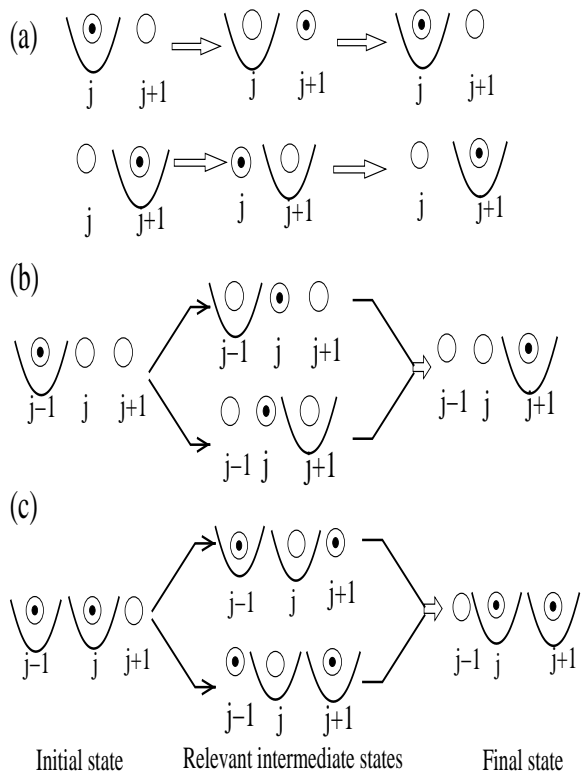}
\caption{{Schematic representation of the hopping processes  
considered in Fig. \ref{hop_fig}
 with intermediate states giving the dominant contributions. 
Here empty circles correspond to empty sites, while circles with small 
dots indicate particle positions. Parabolic curve at a site depicts 
full distortion of the lattice ions at that site with corresponding energy
 $-g^2 \omega_0$ ($+g^2 \omega_0$) if the electron
is present (absent) at that site.
}
}
\label{coeff_fig}
\end{figure}

{
 The coefficients of the last three terms in Eq.(\ref{ai10}) (i.e., $t^{2}/{2g^2\omega_0}$ and
 $t^2 e^{-g^2}/{g^2\omega_0}$),
obtained from second order perturbation theory,
are explained with the help of schematic  diagrams
shown in Fig. \ref{coeff_fig}. We consider two distinct time scales for 
electronic hopping processes between two adjacent sites: 
(i) $\sim t_{eff}^{-1}$  associated with either full distortion 
of the lattice ions to form a small polaronic potential well (of energy $-g^2\omega_0$)
or full relaxation from the small polaronic distortion
and (ii) $\sim t^{-1}$ during which  
negligible distortion/relaxation
of lattice ions occurs.
The upper process in Fig. \ref{coeff_fig}(a) [schematically representing the distortion
for hopping sequence in Fig. \ref{hop_fig}(a)]
shows the process in which an electron goes to its right neighbouring site and 
comes back. The intermediate state 
has energy $+g^2\omega_0$ and corresponds to fully distorted
site $j$ but
without the electron and neighbouring site $j+1$ containing the electron but without 
lattice distortion. 
 To go from the intermediate state to the
final one, the electron at site $j+1$ hops back to the original site $j$ 
without any new distortion taking place. 
Thus, the initial, the intermediate, and the final states all
have identical lattice distortions. 
Hence, the hopping times for both these processes is $t^{-1}$.
Then, from second order perturbation theory,
the numerator of the coefficient is $t^2$ while the denominator (which is energy difference
between the intermediate and the initial states) becomes $2g^2\omega_0$ leading to
the coefficient $\frac{t^2}{2g^2\omega_0}$. The same 
 coefficient results from the process in which an electron goes to its 
left neighbouring site and comes back as depicted by the lower process
in Fig. \ref{coeff_fig}(a). Thus the upper and lower processes in Fig. \ref{coeff_fig}(a)
both yield
the same coefficient $\frac{t^2}{2 g^2\omega_0}$.
}

{
The coefficient of the fourth term $t^2 e^{-g^2}/{g^2\omega_0}$
corresponds to the schematic distortion processes shown in Figs. \ref{coeff_fig}(b) and 
 \ref{coeff_fig}(c) with the pertinent hopping processes being depicted in 
Figs. \ref{hop_fig}(c) and \ref{hop_fig}(d) respectively. 
For the process where one electron at site $j-1$ consecutively hops to its NN site $j$ and then to
NNN site $j+1$, the intermediate states which give
dominant contributions are shown in 
Fig. \ref{coeff_fig}(b). 
After the electron hops from site $j-1$ to site $j$, the upper intermediate state
in Fig. \ref{coeff_fig}(b)  has the same lattice distortion as the initial state.
On the other hand, again in the upper intermediate state
in Fig. \ref{coeff_fig}(b),
when the electron makes the next hop
from site $j$ to $j+1$ to produce the final state, there is a  
 distortion at site $j+1$ with a concomitant relaxation at the initial site $j-1$.
Hence the coefficient contribution, from the hopping process involving the
upper intermediate state
in Fig. \ref{coeff_fig}(b),
becomes $t\times te^{-g^2}/2g^2\omega_0$. 
Alternately, after the electron hops from site $j-1$ to site $j$, 
the initial state may lead to 
the lower intermediate state in Fig. \ref{coeff_fig}(b)
where the distortion at site $j-1$ is completely relaxed
while site $j+1$ gets distorted simultaneously so that the intermediate and the final states
have identical lattice distortions. This process too yields
a coefficient contribution of  $te^{-g^2}\times t/2g^2\omega_0$. The other six possible
intermediate states (corresponding to the remaining possibilities of
full distortion/relaxation at the three sites)
 each give $\sim t^2 e^{-2g^2}$ in the numerator and hence we ignore them.
}

{
 The coefficient
of $\s\s$ or $\sb\s$ pair hopping [depicted by Figs. \ref{hop_fig}(d) and \ref{hop_fig}(e)
respectively]
can be derived from Fig. \ref{coeff_fig}(c) which schematically represents the
dominant contributions. 
The hopping process in Fig. \ref{coeff_fig}(c), with upper (lower) 
intermediate state, represents sequential hopping where an electron at site $j$ hops to $j+1$
and produces the following changes: site $j+1$ is undistorted (distorted), $j-1$ distortion
is unchanged (relaxed),
and $j$ remains distorted. The hopping time for the first hop is $1/t$ ($1/t_{eff}$)
and the change in energy between intermediate and initial states is $2 g^2 \omega_0$.
Next, the electron at site $j-1$ hops to $j$ with site $j-1$ relaxing (remaining undistorted)
and site $j+1$ getting (remaining) distorted. The second hop occurs in time $1/t_{eff}$ ($1/t$).
Thus the $\s\s$ or $\sb\s$ pair hopping process, involving the two dominant intermediate processes 
of Fig. \ref{coeff_fig}(c), yields the coefficient $t^2e^{-g^2}/g^2\omega_0$.
}

\section{Region of validity of our theory}
Our theory has three dimensionless parameters --  Hubbard interaction $U/t$, adiabaticity
$t/\omega_0$, and electron-phonon interaction strength $g$. An increase in $U/t$, while keeping
fixed other dimensionless parameters, will decrease only the spin Heisenberg
AF interaction $J=\frac{4t^2}{U-2g^2\omega_0}$ in the effective Hamiltonian of Eq. (\ref{ai10}).
Furthermore, an increase in $g$ produces an increase in $J$, a decrease in $V$,
and also a decrease in all the various
hopping (i.e., NN hopping, double hopping, and $\sigma\sb$ pair hopping) coefficients in Eq. (\ref{ai10}).

For large values of $g$, the coefficients of NN hopping, NNN  hopping,
$(\sigma\sb)$ pair hopping terms
contain $e^{-g^2}$ and hence are small compared to NN repulsion $V$
 and spin interaction $J$. Thus, unlike the usual $t-J$ model, here the spin-spin interaction
term dominates over the hopping.  Our  model basically describes the competition between spin
interactions and NN repulsion
 with the hopping being a small perturbation. By varying the dimensionless parameters in our model,
the relative importance of $J$ and $V$ terms
can be changed. When the effect of $J$ dominates over that of $V$, we expect formation of an AF cluster.
On the other hand,  when NN repulsion is larger
than spin interaction, the electrons tend to get separated which can lead to different interesting phases.

For our theory to be valid, the following criterion need to be satisfied:
(a) the dimensionless effective Hubbard interaction $U_{eff}/t\gg 1$
so that Eq. $(\ref{ai3})$ can be well approximated by the effective $t-J$ model
given in Eq. $(\ref{ai5})$. Actually, for a one-dimensional Hubbard model,
calculations \cite{avella} show that $U_{eff}/t\gtrsim 8.0$
is sufficient to remove double occupancy;
(b) NN electronic hopping $J_1$ and spin interaction $J$
should both be negligible compared to the phononic energy $\omega_0$ so as to make our
perturbation theory valid;
and (c) the small parameter \cite{sml_parameter1,sml_parameter2} of our perturbation theory $t/(g\omega_0)$
should be kept as small as possible.
Our calculations are done for $t/{g\omega_0}\leq0.5$.

\section{Results and Discussion}
The half-filled case (i.e., average concentration of one electron
per site) is not
very interesting because, owing to exclusion of double occupancy, we have one electron in every site
and consequently no hopping occurs.
Here the system behaves like a
Heisenberg AF chain with NN repulsion having no effect on electronic correlations.
However, at quarter-filling there are enough vacancies for electrons so that,
besides the spin interaction term, NN repulsion
and different kinds of hopping in our model can contribute to a rich phase diagram.
Therefore, we concentrate on the ground state properties at quarter-filling
and analyze the different phases exhibited by our effective Hamiltonian in Eq. (\ref{ai10}) and the
quantum phase transitions (QPT) between these phases for large values of $g$,
i.e., in the strong e-ph coupling limit.

We use a modified Lanczos algorithm \cite{dagatto}
to calculate the ground state.
To get the basis states in the occupation
number representation, we exclude double occupancy and permit only the remaining
three possible electronic states on any site
(i.e., \textuparrow, \textdownarrow, \ and no particle). The size of the resulting Hilbert space prohibits us from calculating
the ground state for large systems. For example, the number of basis states
for 16 sites with 8 electrons (4\textuparrow, 4\textdownarrow) is $^{16}C_8\times ^{8}C_4=900900$ which
is quite large but manageable for calculating the ground state. But
for 20 sites with 10 electrons (5\textuparrow , 5\textdownarrow), the calculation of ground state
requires $^{20}C_{10}\times ^{10}C_5$ basis states which is prohibitive.

We find evidence for several different types of phases depending on parameter values. Because there is
a tendency to form phase separated clusters, various methods are used to identify the phases.
To characterize phase separation, we use a
``$n-$particle normalized clustering probability" parameter NCP(n)
 (defined precisely in Appendix C) which actually measures the normalized probability
of $n-$electron clusters
in the ground state of the system. However, NCP(n) does not tell the number of $\uparrow$ and $\downarrow$
electrons or their arrangement in the cluster. To determine the different phases, we need to employ
additional measures of ordering such as
spin-spin correlation functions, structure factor, etc.

\begin{figure}[t]
\includegraphics[angle=-90,width=0.98\linewidth]{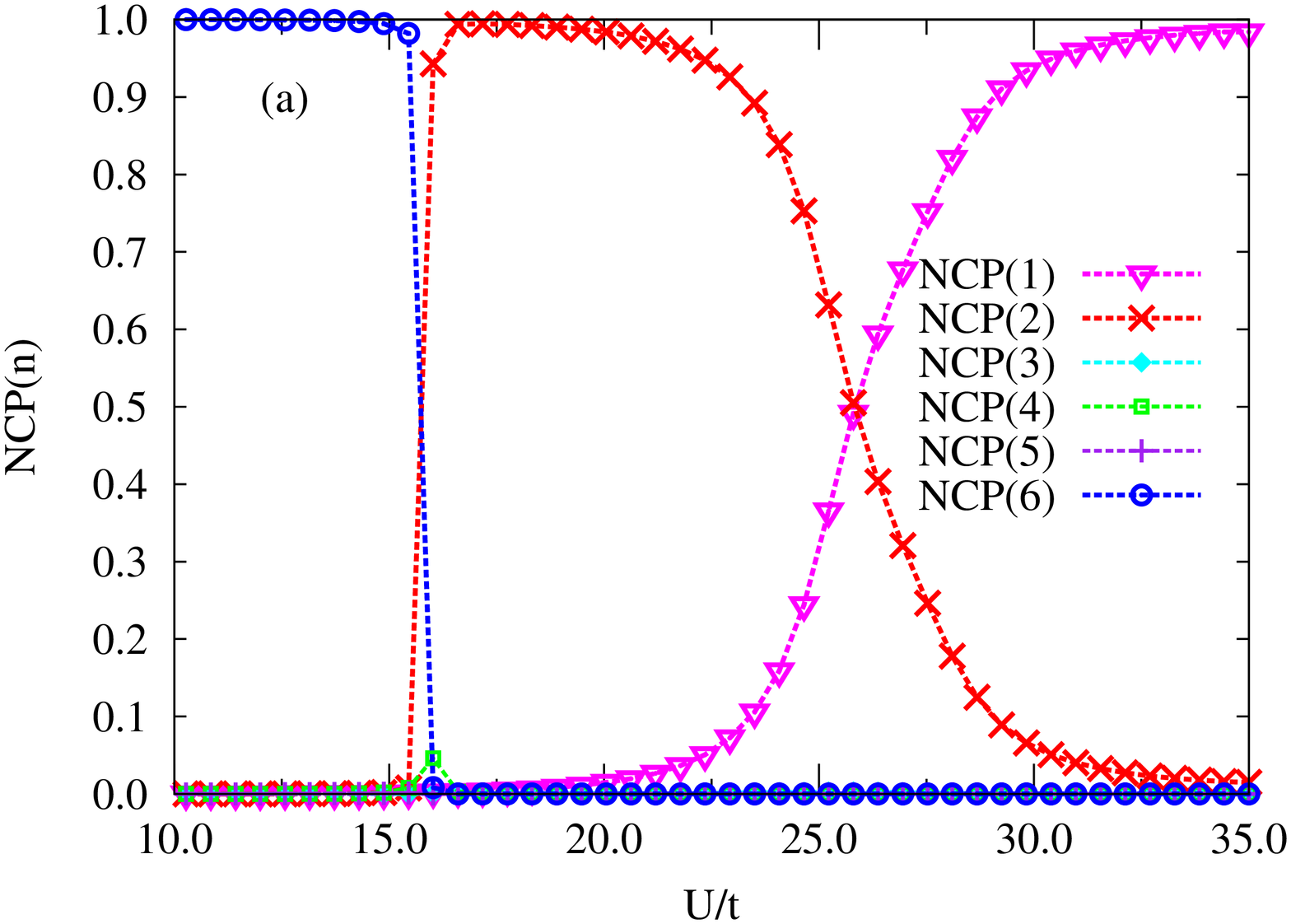}
\includegraphics[angle=-90,width=0.98\linewidth]{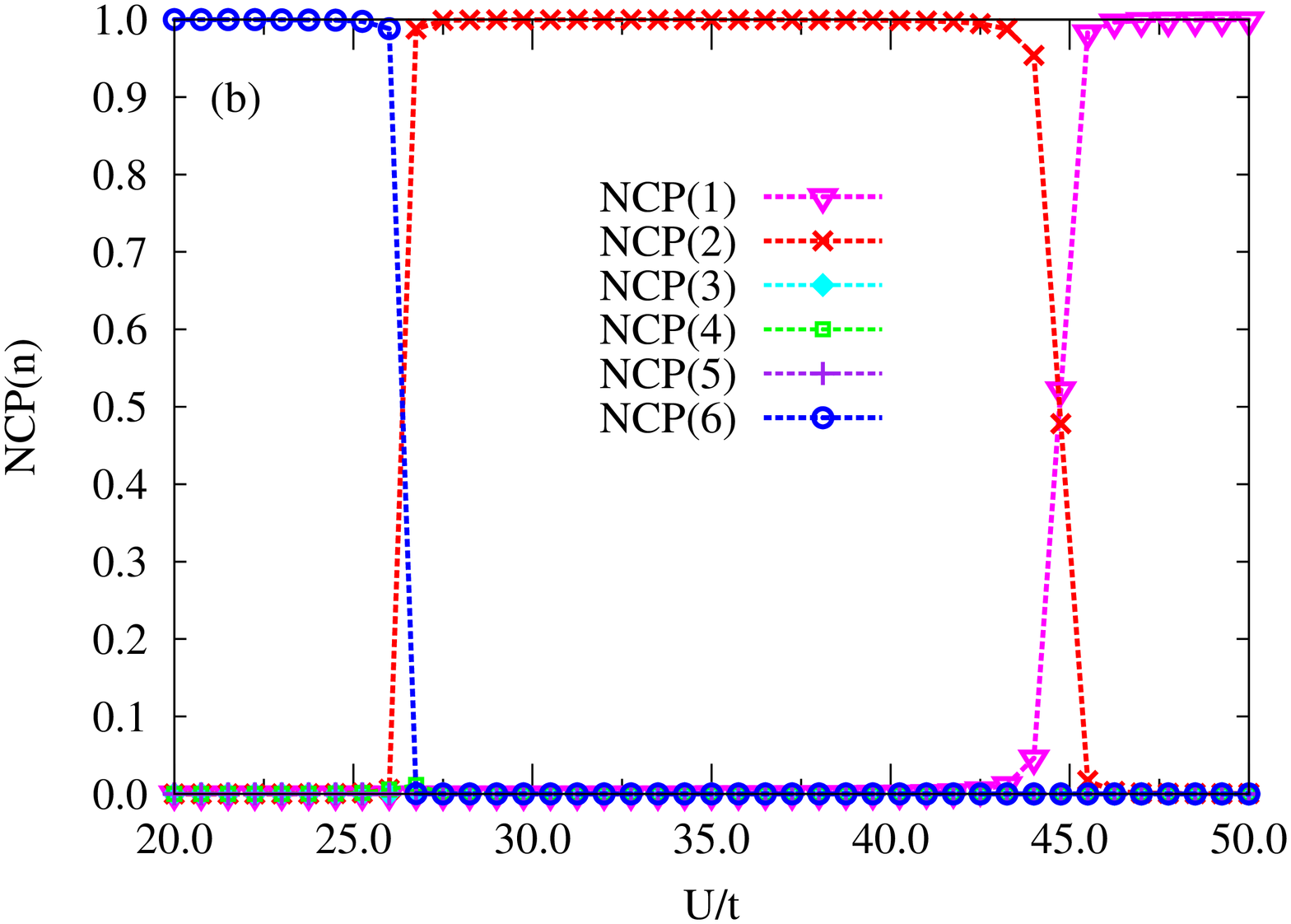}
\caption{(Color online) n-particle normalized clustering probability [NCP(n)] for number of sites =12,
number of electrons = 6 ($3 \uparrow, 3 \downarrow$), $t/\omega_0=1.0$, and
when (a) $g=2.2$ and (b) $g=2.8$. }
\label{NCP}
\end{figure}

\subsection{AF Cluster to Correlated Singlets Transition}
Our numerical calculations were done for $t/\omega_0 = 1.0$ and $t/\omega_0=0.5$.
Since the results are similar for the two values of $t/\omega_0$ considered, here we report only the
calculations for $t/\omega_0=1.0$.
By varying $g$ and $U/t$,
 we change the relative importance of $J$ and $V$ terms and get the different phases of the system.
\begin{figure}[]
\includegraphics[angle=-90,width=0.98\linewidth]{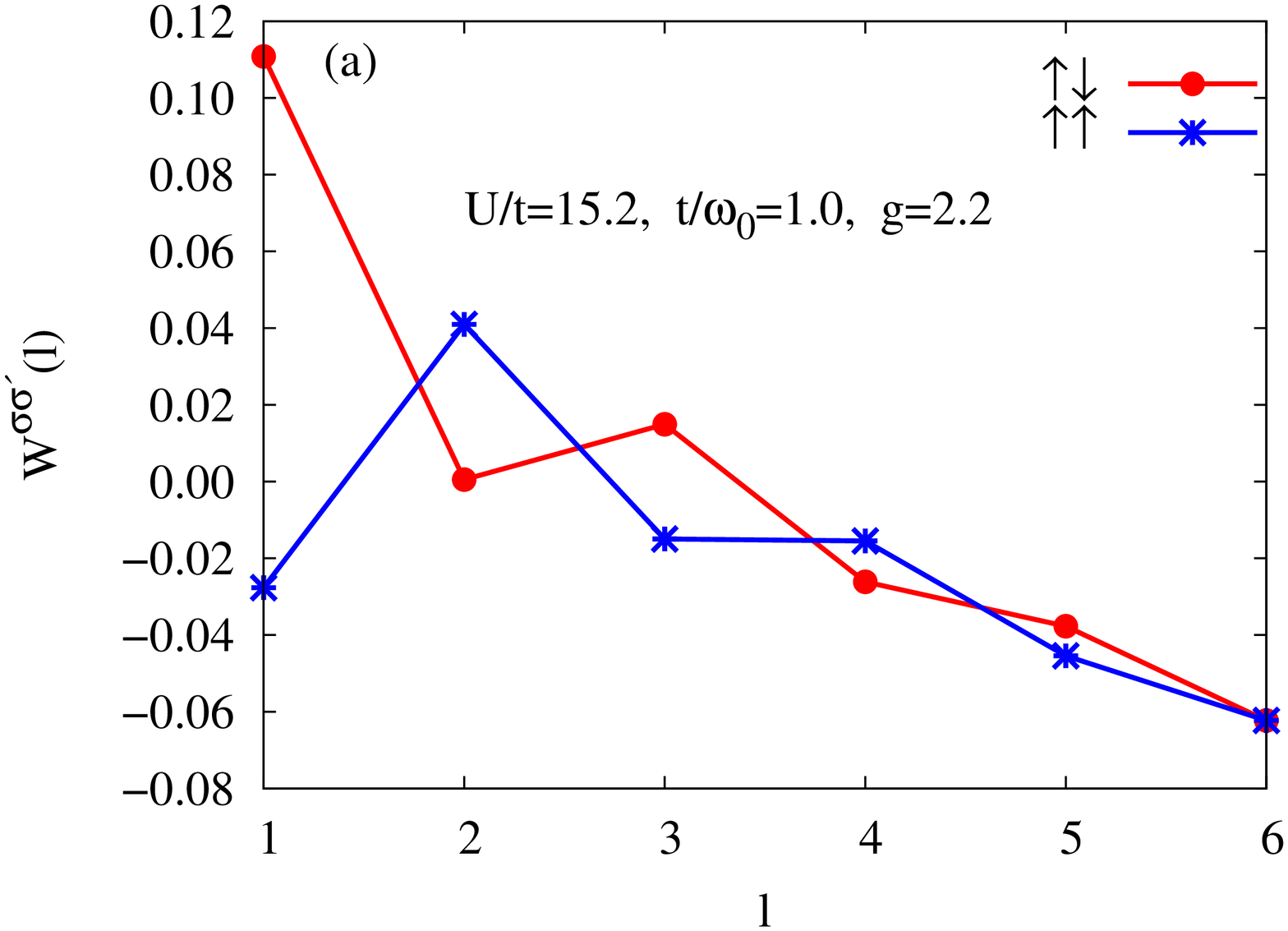}
\includegraphics[angle=-90,width=0.98\linewidth]{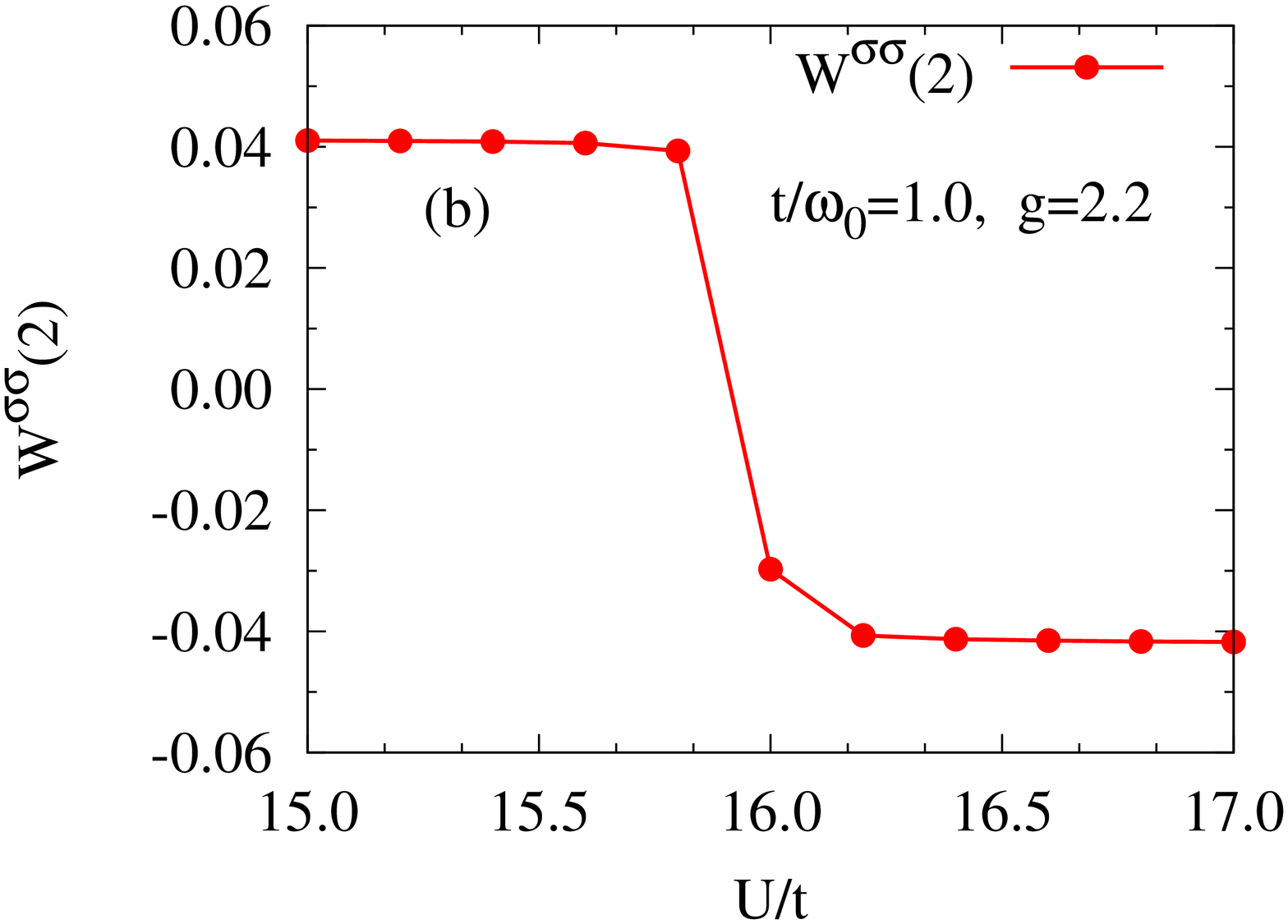}
\caption{(Color online) Plot of the correlation functions (a) $W^{\s\s^{\prime}}(l)$ portraying
an AF cluster for $U/t=15.2$,
$t/\omega_0=1.0$, and $g=2.2$;
and (b) $W^{\s\s}(2)$ versus $U/t$ again for $t/\omega_0=1.0$ and $g=2.2$.}
\label{cf_cluster}
\end{figure}

It is clear from Fig. \ref{NCP}(a) that, for $t/\omega_0=1.0$ and $g=2.2$, for values of $U/t$ up to
about 15.6
the NCP(6) $\approx 1.0$ which implies that
all the particles in the system form a single cluster. This is because the effect of spin interaction
dominates over the NN repulsion.
But upon increasing $U/t$ further, the system undergoes a first-order QPT
with  NCP(6) discontinuously dropping to almost a zero value and
NCP(2) concomitantly jumping abruptly to a value close to $1$. The value of
NCP(2) remains close to 1 up to
$U/t\approx20.0$. Interestingly, for
n=1,3,4,5 particles, NCP(n) remains almost zero up to $U/t\approx23.0$.
For all values of $g$ in the range $2 \leq g \leq 3$, when $U/t$ is increased,
the system transits discontinuously from a single cluster to a phase comprising of pairs of particles.
 This is verified for $g=2.8$ as well in Fig. \ref{NCP}$(b)$.
Moreover, we notice that the correlated pairs phase persists
over a broader window of $U/t$ values when $g$ is larger.
In fact, for all the system sizes considered (i.e., number of sites N = 8, 12, and 16),
we find that the systems manifest a first-order QPT from an AF cluster of size N/2 to a correlated
singlet phase as $U/t$ increases.

Although Fig. \ref{NCP} demonstrates the QPT clearly, it does not give information about the structure/order of the phases.
We will confirm the AF order by analyzing the correlation between density fluctuations of
spins ($\sigma$ and $\s^{\prime}$ separated
by a distance $l$) through
the correlation function
$W^{\s\s^{\prime}}(l)={1\over{N}}\sd_{j}\langle{n_{j}^{\s}n_{j+l}^{\s^{\prime}}}
\rangle-\langle n_{j}^{\s}\rangle\langle n_{j+l}^{\s^{\prime}}\rangle$ with $N$ being the system size.
Here $\langle n_{j}^{\s}\rangle$ is the filling
factor which for quarter-filling is $0.25$. Fig. \ref{cf_cluster}(a) shows the correlation functions
$W^{\s\s^{\prime}}(l)$
for $U/t=15.2$, $t/\omega_0=1.0$, and $g=2.2$, i.e., away from the transition. This clearly shows that in the cluster,
the dominant arrangement of the spins is antiferromagnetic with correlation decreasing with distance $l$.
This is expected because of the transverse spin fluctuation term in the Heisenberg interaction.
Next, in Fig. \ref{cf_cluster}$(b)$,
$W^{\uparrow\uparrow}(2)$ is shown as $U/t$ is varied. There is a sudden jump in $W^{\uparrow\uparrow}(2)$
indicating that the system breaks up into two-particle clusters from an AF cluster.

\begin{figure}[]
\includegraphics[angle=-90,width=0.9\linewidth]{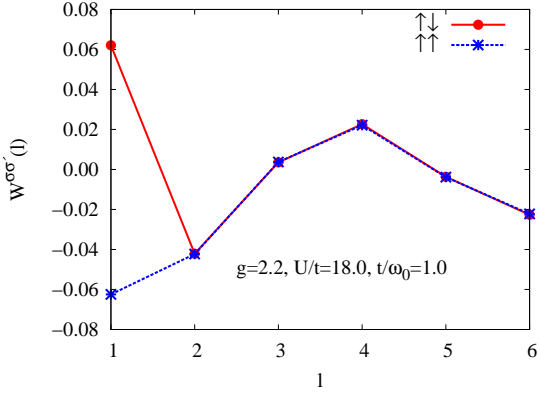}
\caption{(Color online) Plot of correlation functions $W^{\s\s^{\prime}}(l)$ in the correlated
singlet phase for $g= 2.2$, $ U/t=18.0$, and
$t/\omega_0=1.0$.  }
\label{singlet}
\end{figure}

Now, we need to determine the structure of the phase with NCP(2) $\approx 1.0$ in Fig. \ref{NCP}.
The important question is: what is the nature of the pairs?
The answer lies in the
correlation function plot $W^{\s\s^{\prime}}(l)$ shown in Fig. (\ref{singlet}) for $U/t=18.0$ (i.e., deep
 inside the phase).
Here, we see that the value of $W^{\uparrow\uparrow}(1)$ is approximately its minimum possible
value of $-0.0625$ which occurs when
${1\over{N}}\sd_{j}\langle{n_{j}^{\s}n_{j+1}^{\s}}\rangle\approx 0.0$.
This means that each pair is made up of
two opposite spin electrons. Furthermore, $W^{\uparrow\downarrow}(1)=0.0625$ in Fig. (\ref{singlet})
which is expected
because
$W^{\uparrow\downarrow}(1)$ should have a calculated value of $0.25\times0.5-0.0625=0.0625$.
Next, we notice that, for $l\geqslant2$,  $W^{\uparrow\downarrow}(l)$
and $W^{\uparrow\uparrow}(l)$ have the same values.
This means that, given a $\s\sb$ pair, we get not only an $\uparrow\downarrow$ pair at a distance
$l$ with a certain probability but also a
$\downarrow\uparrow$ pair at the same distance and with the same probability.
Thus a pair of opposite spin electrons located at sites $j$ and $j+1$ can either be a singlet or
a triplet with $S^z_{total} =0$. When acted upon by the operator $S_{j}^{+}S_{j+1}^{-}+S_{j}^{-}S_{j+1}^{+}$,
the singlet state yields eigenvalue -1 while the $S^z_{total}=0$ triplet state gives +1.
Thus to know the nature of the spin pairs in the system, we have calculated the quantity
$\sd_{j}\langle(S_{j}^{+}S_{j+1}^{-}+S_{j}^{-}S_{j+1}^{+})\rangle$
 and obtained a value very close to $-3.0$ for $6$ electrons ($3\uparrow$ and $3\downarrow$) for the whole range
of $16.0\lesssim U/t\lesssim 22.5$. Thus the phase is entirely made up of
singlets.  In Fig. \ref{singlet},
 we also notice that  $W^{\uparrow\downarrow}(l)$
and $W^{\uparrow\uparrow}(l)$
 show a peak at $l=4$ and slightly lesser values for $l=3$ and $l=5$ which is indicative of a CDW.
We call this phase a correlated singlet phase. A detailed analysis of this phase at various filling factors
will be presented elsewhere \cite{srsypbl2}.

\begin{figure}[]
\includegraphics[angle=-90,width=0.98\linewidth]{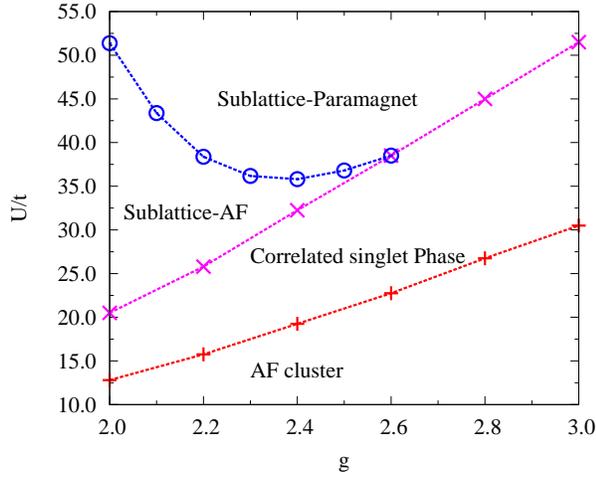}
\caption{(Color online) The phase diagram of the Hubbard-Holstein model at quarter-filling in
 the $t/\omega_0=1.0$ plane.}
\label{phase_diag}
\end{figure}

The transition from an AF cluster to correlated singlets can be explained by invoking Bethe ansatz results.
Suppose we have a Hamiltonian of the type $H=\tilde J\sd_{j}S_j.S_{j+1}+\tilde V\sd_{j}n_jn_{j+1}$.
Then, for the cluster regime,
Bethe ansatz yields energy/site $=-0.443 \tilde J+2\frac{\tilde V}{2}$ where as for separated singlets in
the correlated singlet phase the energy/site $=-\tilde J(0.375)+\frac{\tilde V}{2}$. Thus the cluster regime
prevails when
$-0.443 \tilde J+\tilde V\leq-\tilde J(0.375)+\frac{\tilde V}{2}$, i.e., $\tilde V\leq 0.136\tilde J$.
Now, if in Eq. (\ref{ai10}) we neglect all the hopping terms (which is true for large $g$), based
on the above analysis, we
obtain the condition for existence of the
cluster phase to be $(2V-\frac{J}{4})\lesssim 0.136 J$ or equivalently
$U/t\lesssim 3.544g^2\omega_0/t$. This condition gives a very good estimation of the actual phase boundary in
the phase diagram
given in Fig. \ref{phase_diag}.
\subsection{Transition from correlated singlets to spins in one sublattice}
Here we discuss how, for a fixed $g$ and larger values of $U/t$, the correlated
singlets break up into spins
occupying only one sublattice . It is clear
from Fig. \ref{NCP}(a) that this is a second-order phase transition at smaller values of $g$ (such as $g=2.2$)
because NCP(1) [NCP(2)]
increases [decreases] continuously as $U/t$ is raised. On the other hand, this transition tends
towards first-order at larger values of $g$ [say $g=2.8$ as shown in Fig. \ref{NCP}(b)]. This
transition can be explained as follows.
As $U/t$ is increased, $J$ gets reduced and NN repulsion
starts dominating over spin interaction leading to the break up of singlets and
single spins separating out to occupy one sublattice only.
\begin{figure}[]
\includegraphics[angle=-90,width=0.98\linewidth]{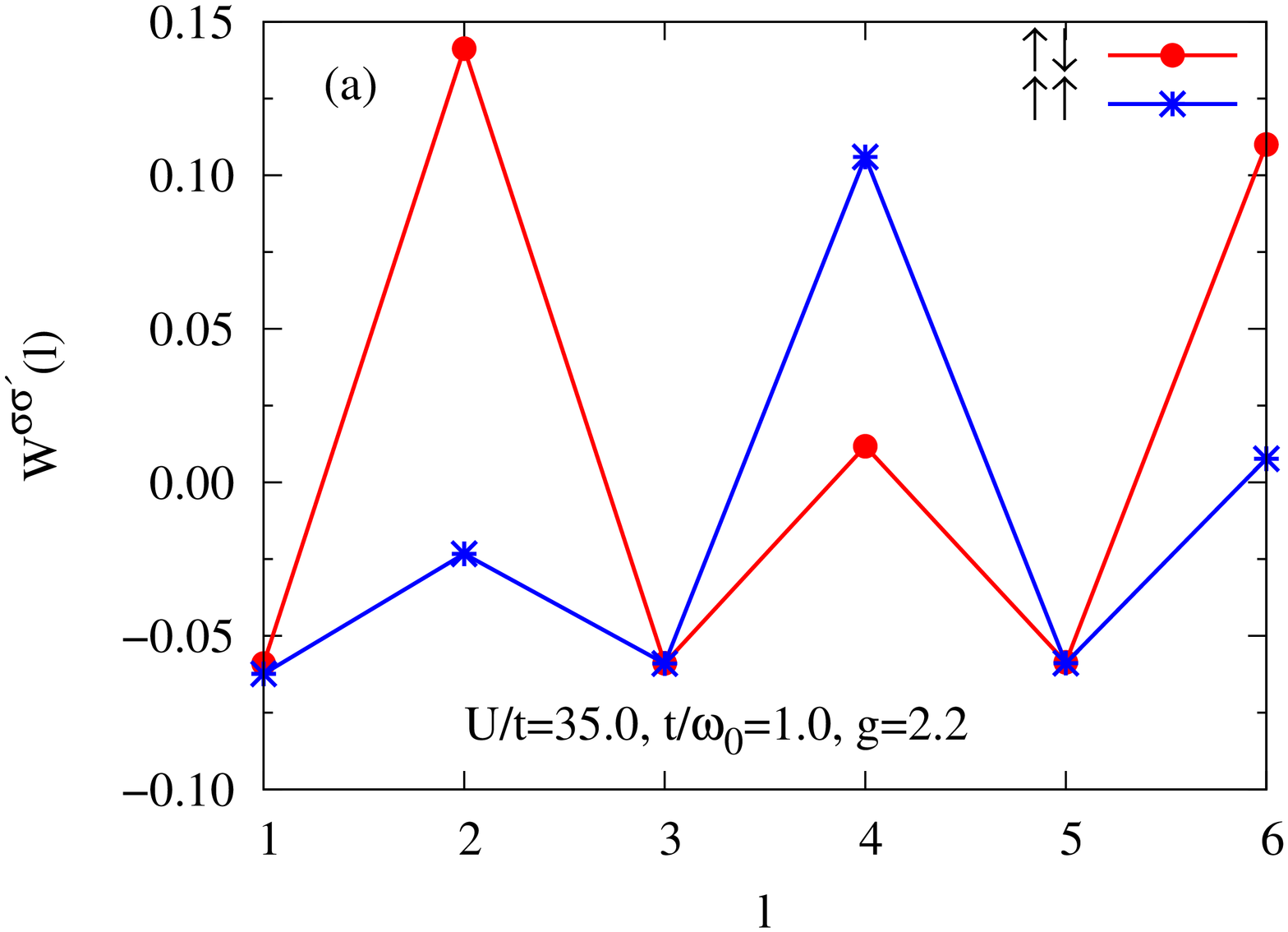}
\includegraphics[angle=-90,width=0.98\linewidth]{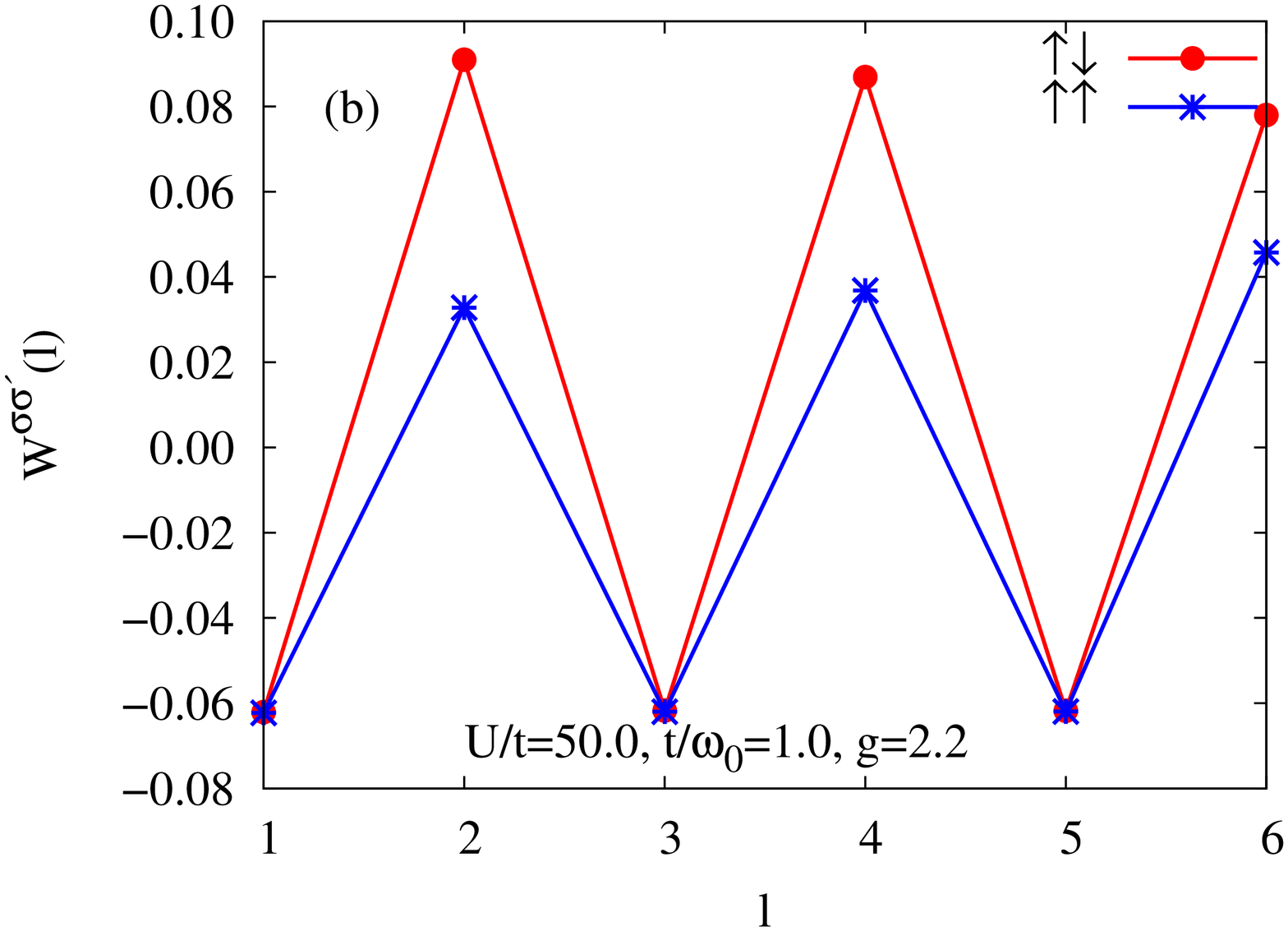}
\caption{(Color online) Correlation functions $W^{\s\s^{\prime}}(l)$ at $t/\omega_0=1.0$, $g=2.2$, and
 for (a) $U/t=35.0$ depicting AF order in one sublattice;
and for (b) $U/t=50.0$ displaying paramagnetic phase in one sublattice.}
\label{cf_pair2AF}
\end{figure}

\begin{figure}[]
\includegraphics[angle=-90,width=0.98\linewidth]{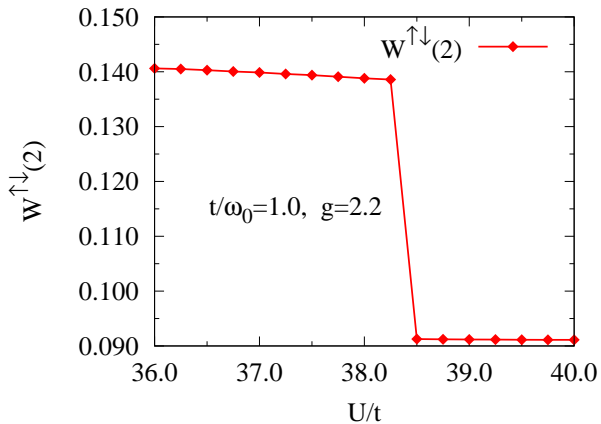}
\caption{(Color online) Plot of the correlation function $ W^{\uparrow\downarrow}(2)$ versus $U/t$.}
\label{AF_para}
\end{figure}

\begin{figure}[]
\includegraphics[angle=-90,width=0.98\linewidth]{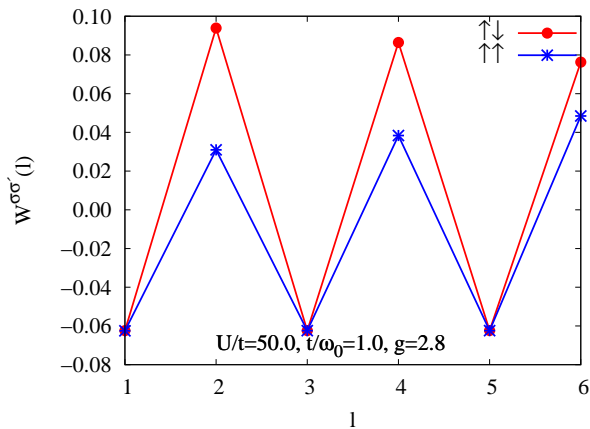}
\caption{(Color online)  Correlation functions $W^{\s\s^{\prime}}(l)$
depicting spin disorder in one sublattice for $U/t=50.0$, $t/\omega_0 = 1$, and $g=2.8$.}
\label{cf_singles}
\end{figure}

\subsubsection{Transitions at smaller $g$}
For smaller values of $g$ (i.e., $2<g<2.6$), the QPT is not so sharp. In fact,
for $g=2.2$ [as shown in Fig. \ref{NCP}(a)],
the singlet pairs and separated single spins coexist in the range $23 < U/t < 30$.
But when $U/t$ is
increased further, almost all the pairs are broken and the electrons occupy one sublattice only. The
correlation functions $W^{\uparrow\downarrow}(l)$ and $W^{\uparrow\uparrow}(l)$, depicted
in Fig. \ref{cf_pair2AF}$(a)$
for $U/t=35.0$ and $g=2.2$, show that the dominant structure has AF order in one sublattice
(i.e, the spins arrange themselves as $\uparrow-\downarrow-\uparrow-\downarrow-...$).
This can be surmised from the fact that the correlation functions
$W^{\uparrow\downarrow}(l)$ and $W^{\uparrow\uparrow}(l)$ have dominant peaks at alternate even values of $l$.
 This spin arrangement is preferred by the system because it can gain
energy due to AF interaction $(J)$ by virtual hopping to NN site and returning back.
For very large values of $U/t$, the
spin interaction strength $J$ becomes negligible and the electrons do not shown any spin order
but still occupy only one sublattice. For this spin disordered situation, the values of
$W^{\uparrow\downarrow}(l)$ and $W^{\uparrow\uparrow}(l)$ for the system under consideration
[i.e., $12$ sites with $6$ electrons ($3\uparrow,3\downarrow$)] can be calculated as follows.
The probability of getting one $\uparrow$ spin at any site, in a $12$-site system, is $0.25$.
 Now, as the spins are residing in only one sublattice,
the probabilities of finding the remaining $3 \downarrow$ and $2 \uparrow$ electrons, in
any of the remaining $5$ sites of the same sublattice, are
$3/5$ and $2/5$ respectively. Hence, $W^{\uparrow\downarrow}(l)=0.25(3/5)-0.0625=0.0875$ and
$W^{\uparrow\uparrow}(l)=0.25(2/5)-0.0625=0.0375$ for $l=2,4,6,8,10$.
The above predicted values of the correlation functions
$W^{\uparrow\downarrow}(l)$ and $W^{\uparrow\uparrow}(l)$
 match well with the calculated values of these functions
for $U/t=50.0,g=2.2$ as shown in Fig. \ref{cf_pair2AF}(b).
Another interesting point is that there is a first-order phase transition from AF order
to paramagnetic order in one
sublattice of the system as can be seen from the sudden jump (when $U/t$ is varied)
in the correlation function $W^{\uparrow\downarrow}(2)$
depicted in Fig. \ref{AF_para}.

\subsubsection{Transition for larger $g$ values}
 For $g\gtrsim 2.6$, when $U/t$ is increased,
the correlated singlet phase transits to a  paramagnetic state (in one sublattice) directly unlike
the earlier case for smaller values of $g$ such as $g=2.2$
(see Fig. \ref{cf_singles}). This transition is of first-order character
 as shown in Fig.\ref{NCP}$(b)$ where  NCP(1) and
 NCP(2) show simultaneous sharp jumps in opposing directions.

\section{The Phase Diagram}
The phase diagram of the Hubbard-Holstein model at quarter-filling in a twelve-site system
for the $t/\omega_0=1.0$ plane is
shown in Fig. $\ref{phase_diag}$. This is obtained by producing plots of  NCP(n) versus $U/t$
for various values of $g$. Two such plots are shown in Fig. $(\ref{NCP})$ for
values of $g=2.2$ and $g=2.8$. Then, from the NCP plots, we find phase transition points in the $g-U/t$ plane to obtain
the phase diagram. Furthermore, using selective choice of parameter values, we also found that
the qualitative features of the phase diagram remained the same
 even for the larger sixteen-site system.

In the phase diagram shown in Fig. \ref{phase_diag}, the system
transits, for all values of $g$ in the range $2 \leq g \leq 3$, from an AF cluster to a correlated singlet phase
discontinuously when $U/t$ is increased.
 For $ g \lesssim 2.6$, further increase in $U/t$ drives the system continuously
from a correlated
singlet phase to a single-sublattice AF phase.
Next, at even higher values of $U/t$ and again for $g \lesssim 2.6$, the system jumps discontinuously from an
AF order to a paramagnetic
 order in a single sublattice. Contrastingly, for $g \gtrsim 2.6$,
the system transforms directly from a correlated singlet phase to a paramagnetic phase by a close-to-discontinuous jump.
\section{Conclusions}
By analyzing the probability of occurrence of different cluster sizes and by studying various correlation functions
that result from our effective Hubbard-Holstein Hamiltonian,
we deduced the ground state phase diagram (Fig. \ref{phase_diag}) at quarter-filling and in the non-adiabatic
regime. In particular we find that strong electron-phonon coupling stabilizes a 
correlated singlet phase that is charge-density-wave like, which would be absent 
in the pure Hubbard model. Notice that this phase is also quite distinct from a 
Peierls-like ( ``bond-order'') wave, being driven by onsite correlations alone.
Our analysis and results
should be of relevance to systems such as fullerides, lower-dimensional organic conductors,
magnetic oxides, interfaces between oxides, where there is evidence for both strong Coulomb correlation as well as electron-phonon coupling.
In higher dimensions,
the concurrent NN spin-spin and NN repulsion interactions, could lead to even richer physics
such as competition between or coexistence of charge ordering and superconductivity, non-Fermi liquid
behavior, etc.

\appendix
\section{}
In this appendix, we will demonstrate the validity of Eq. (\ref{ai8}).
Let us assume a Hamiltonian of the form $H=H_0+H_1$ where
 the eigenstates of $H_0$ are all separable and are of the form
$|n,m\rangle=|n\rangle_{el}\otimes|m\rangle_{ph}$ and $H_1$ is the electron-phonon
interaction perturbation term of the form given in Eq. (\ref{ai7.2}).\\

   After a canonical transformation, we get
\be
\tilde H&=&e^{S}He^{-S}\nonumber\\
&=& \!\!H_0\!+\!H_1\!+\![H_0+H_1,S]\!\!+\!\frac{1}{2}\left[[H_0\!+\!H_1,S],S\right] .
\label{ap1}
\ee
In the ground state energy, we know that the first-order perturbation term is zero.
To eliminate the first-order term in $H_1$, we let $H_1+[H_0,S]=0$. Consequently, we obtain
\be
\langle n_1,m_1|S|n_2,m_2\rangle=-\frac{\langle n_1,m_1|
H_1|n_2,m_2\rangle}{(E_{n_1,m_1}-E_{n_2,m_2})} .
\label{ap2}
\ee
We now assume that both the exponentially reduced electronic NN hopping energy $te^{-g^2}$ and
the Heisenberg spin interaction energy $J$ are negligible
compared to the phononic energy $\omega_{0}$ which is true
for large values of $g$. This implies that
$(E_{n_1,m_1}-E_{n_2,m_2}) \simeq (E_{m_1}^{ph}-E_{m_2}^{ph})$.
Then, Eq. (\ref{ap2}) simplifies to the form
\be
{_{ph}\!\langle m_1| S|m_2\rangle_{ph}} = -\frac{{_{ph}\!\langle m_1|
H_1|m_2\rangle_{ph}}}{(E_{m_1}^{ph}-E_{m_2}^{ph})} .
\label{ap2.2}
\ee
Now, using Eqs. (\ref{ap1}) and (\ref{ap2.2}), we obtain
\begin{widetext}
\be
{_{ph}\!\langle m_1|\tilde H|m_2\rangle_{ph}}  & \simeq & {_{ph}\!\langle m_1|H_0|m_2\rangle_{ph}}
+\frac{1}{2}\sum_{\bar m} \left[~ {_{ph}\!\langle m_1|H_1|\bar m\rangle_{ph}}~
{_{ph}\!\langle \bar m|S|m_2\rangle_{ph}}
-{_{ph}\!\langle m_1|S|\bar m\rangle_{ph}}~{_{ph}\!\langle\bar m|H_1|m_2\rangle_{ph}} \right]
\nonumber\\
&\simeq& {_{ph}\!\langle m_1|H_0|m_2\rangle_{ph}}+\frac{1}{2}\sum_{\bar m}
{{_{ph}\!\langle m_1|H_1|\bar m\rangle_{ph}}~
{_{ph}\!\langle \bar m|H_1|m_2\rangle_{ph}}}
\left [ \frac{1}{E_{m_2}^{ph}-E_{\bar m}^{ph}}
+\frac{1}{E_{m_1}^{ph}-E_{\bar m}^{ph}} \right ]
.
\label{ap3}
\ee
\end{widetext}

\begin{widetext}
\section{Projection onto singly occupied subspace}
\label{proj_singly}
In this appendix we evaluate each term in Eq. (\ref{ai9}) by projecting out
double occupancy. Every fermionic operator $c_{j\sigma}$ is subject to the
 replacement$c_{j\sigma}\rightarrow c_{j\sigma} (1-n_{j\sb})$ so as to incorporate the action of the
single subspace projection operator $P_s$. In the first two terms of Eq. (\ref{ai9}),
the action of $P_s$ implies that spin index $\s=\s{'}$.
Thus, these first two terms 
correspond to the process
where an electron hops to a neighboring site and comes back. 
Then, the projected form of the 
first term is evaluated to be
\be
P_s\left(c_{j+1\s}^{\dagger}c_{j\s}P_{s}^2c_{j\s}^{\dagger}c_{j+1\s}\right)P_s&=&(1-n_{j+1\sb})
c_{j+1\s}^{\dagger}c_{j\s}(1- n_{j\sb})(1-n_{j\sb})c_{j\s}^{\dagger}c_{j+1\s}(1-n_{j+1\sb})\nonumber\\
&=& (1-n_{j+1\sb})c_{j+1\s}^{\dagger}c_{j\s}c_{j\s}^{\dagger}c_{j+1\s}(1-n_{j\sb}) .
\label{apb1}
\ee

Similarly, for the second term we get
\be
P_{s}c_{j\s}^{\dagger}c_{j+1\s}P_{s}^2c_{j+1\s}^{\dagger}c_{j\s}P_{s}=(1\!-\sp n_{j+1\sb})c_{j\s}^{\dagger}
c_{j+1\s}c_{j+1\s}^{\dagger}c_{j\s}(1\!-\sp n_{j\sb}) .
\label{apb2}
\ee
The above two processes are depicted in Figs. \ref{hop_fig}(b) and \ref{hop_fig}(a) respectively
and add up to
\be
&&P_sc_{j+1\s}^{\dagger}c_{j\s}P_{s}^2c_{j\s}^{\dagger}c_{j+1\s}P_s+P_{s}c_{j\s}^{\dagger}c_{j+1\s}P_{s}^2c_{j+1\s}^
{\dagger}c_{j\s}P_{s}
=(1\!-\sp n_{j+1\sb})(1\!-\sp n_{j\sb})(n_{j\s}-n_{j+1\s})^2 .
\label{apb3}
\ee

Next, let us consider the third and fourth terms in Eq. (\ref{ai9}). 
These two terms depict the process of an electron
hopping to its neighboring site and then to the next-neighboring site.
The process described by the fourth term is shown in Fig. \ref{hop_fig}(c) 
and the third term is hermitian conjugate to it.
Obviously, both neighboring and
next-neighboring sites should be empty and here too we should take $\s=\s{'}$. 
On using the projection operator $P_{s}$,
these two terms yield
\be
&&P_{s}c_{j-1\s}^{\dagger}c_{j\s}P_{s}^2c_{j\s}^{\dagger}c_{j+1\s}P_{s}+P_{s}c_{j+1\s}^{\dagger}c_{j\s}P_{s}^2
c_{j\s}^{\dagger}c_{j-1\s}P_{s}\nonumber\\
&=&(1-n_{j+1\sb})(1-n_{j\sb})(1-n_{j-1\sb})\left[c_{j+1\s}^\d(1-n_{j\s}) c_{j-1\s}+{\rm H.c.}\right] .
\label{apb4}
\ee

Finally, we will consider the last two terms in Eq. (\ref{ai9}). These two terms represent the
consecutive hopping processes
 where an electron of spin $\sigma$ at site $j$ hops to its neighboring site
 $j+1$ $(j-1)$ followed by
 another electron of spin $\s{'}$
 at site $j-1$ $(j+1)$ hopping to site $j$. This successive hopping
process  may be termed $\s{'}\s$ pair hopping. Here, the spin indices $\s$ and $\s{'}$ can be same or different.

Now, the fifth term can be decomposed into two terms as follows of which :
\be
&&P_{s}c_{j\s}^{\dagger}c_{j+1\s}P_{s}^2c_{j-1\sb}^{\dagger}c_{j\sb}P_{s}+P_{s}c_{j\s}^{\dagger}c_{j+1\s}
P_{s}^2c_{j-1\s}^{\dagger}c_{j\s}P_{s}\nonumber\\
&=&(1-n_{j+1\sb})(1-n_{j-1\s})c_{j\s}^{\d}c_{j+1\s}c_{j-1\sb}^\d c_{j\sb}
+(1-n_{j+1\sb})(1-n_{j\sb})(1-n_{j-1\sb})c_{j-1\s}^\d(-n_{j\s}) c_{j+1\s}\label{apb5}
\ee
The sixth term is the hermitian conjugate of the fifth term.
Thus, upon adding the last two terms in Eq. (\ref{ai9}), we arrive at the expression
\be
&&\sp(1-n_{j+1\sb})(1-n_{j-1\s})\left[c_{j\s}^{\d}c_{j+1\s}c_{j-1\sb}^\d c_{j\sb}+{\rm H.c.}\right]\nonumber\\
&&+(1-n_{j+1\sb})(1-n_{j\sb})(1-n_{j-1\sb})\left[c_{j+1\s}^\d(-n_{j\s}) c_{j-1\s}+{\rm H.c.}\right]\label{apb20}
\ee

The second term in Eq. (\ref{apb20}) corresponds to the process depicted 
in Fig. $\ref{hop_fig}(d)$
and the hermitian conjugate part of the first term is indicated by Fig. $\ref{hop_fig}(e)$.  
\end{widetext}

\section{Definition of n-particle normalized clustering probability [NCP(n)]}
We are dealing with a system of $N$ sites containing an even number $N/2$ electrons. The
ground state has equal number of $\uparrow$ and $\downarrow$
spin electrons. We get the basis states $(\phi_i)$ in occupation number representation
by populating $N$ sites with $N/2$ electrons using all possible combinations
with the constraint that each site can have only $3$ possibilities
(i.e., \textuparrow , \textdownarrow, \ and no particle) with double occupancy being excluded.
 The ground state $|\psi_0 \rangle$ is obtained as a
linear combination of these basis states: $|\psi_0\rangle=\sd_i a_i\phi_i$ where $a_i$ are
the probability amplitudes. Then, we calculate the $n-$particle Normalized Clustering Probability, i.e.,
 NCP(n) from the ground state using the following procedure:\\
\begin{enumerate}
\item Initialize the clustering probability ${\rm CP(n)} =0.0$ for $n=1$ to $N/2$.
\item Consider a basis state $\phi_i$ with a corresponding coefficient $a_i$.
\item Find an empty site (say $j$) in the basis state $\phi_i$.
\item Start searching sequentially from site $j+1$ onwards for occupied sites with index larger than $j$.
\item Count the number of occupied sites until another empty site, say $k(>j)$, is reached. Since
the size $n$ of the unbroken cluster of electrons
between the two empty sites $j$ and $k$ is given by $n=k-j-1$, add $a^2_{i}$ to ${\rm CP(k-j-1)}$.
Then, again start searching for electrons from site
$k+1$ onwards until the next empty site is reached and again obtain the next unbroken cluster size $n_1$.
Similar to the previous case, add
$a^2_{i}$ to ${\rm CP(n_1)}$. Continue the searching process for the whole system, i.e., from site
$j$ to site $(j+N)$ with site $(j+N)$ being
 equivalent to site $j$.
\item Repeat steps $3$ to $5$ successively for all the basis states $\phi_i$ with corresponding coefficients
$a_i$ to get
$\rm{CP(n)}$ where $1\leqslant n\leqslant N/2$.
\item Finally, by normalization, calculate NCP for $n-$particle cluster
using the expression $\rm{NCP(n)=CP(n)/{\sd_{n=1}^{N/2}CP(n)}}$ for $1\leqslant n\leqslant N/2$.
\end{enumerate}

\end{document}